\documentclass[a4paper,11pt]{article}
\usepackage{pos}

\def\del{\partial}

\def\sst{\scriptscriptstyle}

\long\def\comment#1{ }

\newcommand\redout{\bgroup\markoverwith{\textcolor{red}{\rule[.5ex]{2pt}{2pt}}}\ULon}

\def\0{{\boldsymbol 0}}

\def\l{{\boldsymbol l}}

\def\and{\qquad\text{and}\qquad}

\def\tform{{t_\text{f}}}
\def\tdecoh{t_\text{d}}

\def\jet{\text{jet}}

\def\pT{p_{\sst T}}

\newcommand{\beq}{\begin{eqnarray}}
\newcommand{\eeq}{\end{eqnarray}}
\newcommand{\be}{\begin{eqnarray*}}
\newcommand{\ee}{\end{eqnarray*}}
\newcommand{\bal}{\begin{align}}
\newcommand{\eal}{\end{align}}
\newcommand{\rmd}{{\rm d}}
\newcommand{\dd}{{\rm d}}

\title{Jet suppression and azimuthal anisotropy at RHIC and LHC}

\author[a]{Yacine Mehtar-Tani}
\author[b,c,d]{Daniel Pablos}
\author*[e]{Konrad Tywoniuk}

\affiliation[a]{Physics Department, Brookhaven National Laboratory,\\
Upton, NY 11973, USA}

\affiliation[b]{INFN, Sezione di Torino, via Pietro Giuria 1, I-10125 Torino, Italy}
\affiliation[c]{Departamento de F\'isica, Universidad de Oviedo, Avda. Federico Garc\'ia Lorca 18, 33007 Oviedo, Spain}
\affiliation[d]{Instituto Universitario de Ciencias y Tecnolog\'ias Espaciales de Asturias (ICTEA), Calle de la Independencia 13, 33004 Oviedo, Spain}

\affiliation[e]{Department of Physics and Technology, University of Bergen,\\
All\'egaten 55, 5007 Bergen, Norway}

\emailAdd{mehtartani@bnl.gov}
\emailAdd{daniel.pablos.alfonso@to.infn.it}
\emailAdd{konrad.tywoniuk@uib.no}

\abstract{
Jets are multi-partonic systems that develop before interactions with the quark-gluon plasma set in and lead to energy loss and modifications of their substructure. Jet modification depends on the degree to which the medium can resolve the internal jet structure that is dictated by the physics of coherence governed by a critical angle $\theta_c$. Using resummed quenching weights that incorporate the IOE framework for medium-induced radiation and embedding the system into a realistic heavy-ion environment we compute the dependence of jet suppression on the cone angle $R$ of the jet, both at RHIC and the LHC. At RHIC kinematics we see a very mild cone angle dependence for the range of $R$ studied, similar to what was found at the LHC. We also present results for the jet azimuthal anisotropy $v_2$ as a function of $R$. We observe that as centrality is decreased, $v_2$ for moderate $R$ jets sequentially collapse towards the result for small $R=0.1$. The reason of this sequential grouping is the evolution of $\theta_c$ with centrality due to its strong dependence on the in-medium traversed length. For jets with $R> \theta_c$, traversing shorter lengths within the medium will make a larger difference than for jets with $R<\theta_c$, since the size of the resolved phase-space over which quenching weights are resummed will be reduced. For this reason, $v_2(R)$ is quite sensitive to the typical value of $\theta_c$ at a given centrality.
}

\FullConference{HardProbes2023\\
 26-31 March 2023\\
 Aschaffenburg, Germany\\}

\begin{document}
\maketitle

\section{Introduction}

Hard probes such as high-$p_T$ identified hadrons or jets have been one of the most prominent observables carrying the imprint of the hot and dense nuclear medium created in ultra-relativistic heavy-ion collisions. Jet yield suppression, commonly expressed in terms of the so-called nuclear modification factor $R_{AA}(p_T)$, represents the most paradigmatic example of the phenomenon of jet quenching, or how energetic colored objects get modified due to their interaction with deconfined QCD matter. A second essential observable is jet azimuthal anisotropy which, roughly speaking, is sensitive to path-length differences in jet suppression due to the relative orientation in the transverse plane of the jet direction $\phi$ with respect to the event plane of the collision $\Psi_R$. These observables have established a consistent picture of the quark-gluon plasma (QGP) as an opaque medium to hadronic probes and highlighted the role of the geometry of the heavy-ion collisions as an important factor that modulates the strength of the interaction.

From the point of view of perturbative QCD, there is a crucial difference between high-$p_T$ hadrons and fully reconstructed jets. While the latter involves the fragmentation of a leading fragment originating from the hard matrix element, jets are multi-scale probes which are sensitive to soft and collinear radiation within the permitted phase space as given by the jet total $p_T$ and cone angle $R$. The differences in radiation patterns imply that these two observables should be treated differently in a medium as well. In particular, the differences in the fragmentation process will directly affect how these observables become sensitive to the opaque nature of the QGP.

The theory of how highly energetic quarks or gluons interact with a deconfined background is well established for both elastic and inelastic processes.\footnote{For a strongly coupled medium, the gauge-gravity duality offers insights about the drag forces suffered by the projectile.} Focussing on the latter, in a sufficiently dense medium, where the size of the medium $L$ becomes significantly bigger than the mean free path $\lambda_{\rm mfp}$, i.e. $L \gg \lambda_{\rm mfp}$, multiple scattering with the medium leads to the frequent emission of soft quanta that subsequently rapidly cascade further down to the medium temperature scale and spread out over large angles. This turns out to be an efficient mechanism for \emph{energy-loss}, i.e. the transport of energy from the leading parton to soft modes at large angles \cite{Blaizot:2013hx}.

The modifications of a leading fragment, which predominantly contributes to the high-$p_T$ inclusive hadron spectrum, is closely related to how a single parton is affected by medium interactions. How to arrive at the modifications of a jet is a wholly different question. Being a multi-partonic object, a reconstructed jet should be sensitive to the energy-loss suffered by several of its constituents and could experience modifications to its internal structure by capturing genuine medium-induced bremsstrahlung or medium recoil within its cone angle. A new framework for calculating energy-loss of jet observables has recently been developed \cite{Mehtar-Tani:2017ypq,Mehtar-Tani:2017web,Mehtar-Tani:2021fud}, see also \cite{Caucal:2021cfb} for calculations of jet substructure observables. In particular, one has identified the relevant scales where medium-modifications set it. Furthermore, the physics of QCD coherence also unambiguously identifies a minimal critical angle $\theta_c$ which the medium is not able to resolve. 

In a previous publication \cite{Mehtar-Tani:2021fud}, we have calculated the $R$-dependent jet spectrum in heavy-ion collisions at LHC, obtaining excellent agreement with experimental data as a function of $p_T$, the cone angle $R$ and the centrality of the collisions. In these proceedings, we extend the analysis to the azimuthal anisotropy through the harmonic coefficient $v_2(p_T,R)$ as a function of the same variables. The inherent sensitivity of $v_2$ to path-length differences in the azimuthal plane makes it also an excellent measure of coherence physics.

\section{Methodology for jets in heavy-ion collisions}

\begin{figure}[t!]
\centering
\includegraphics[width=0.41\textwidth]{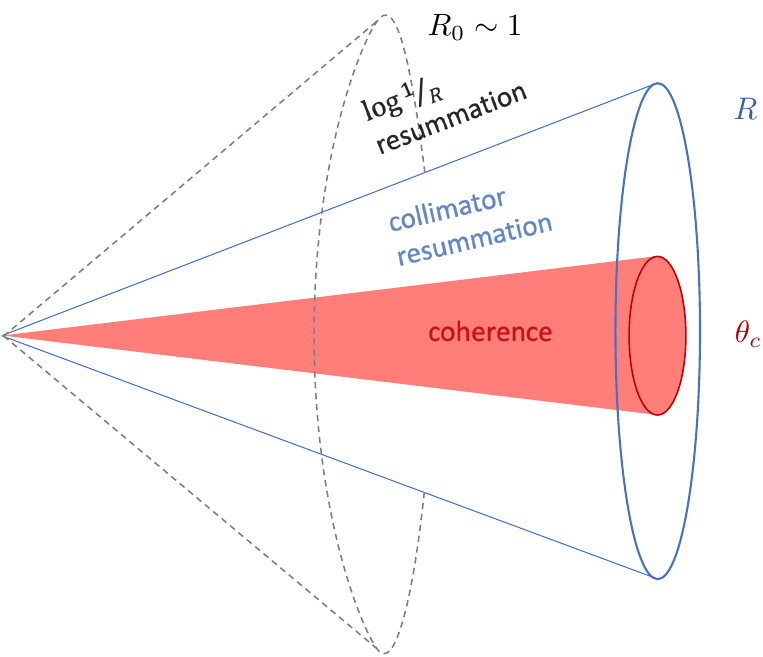}
\caption{Depiction of the two resummation schemes needed to compute the jet spectrum in heavy-ion collisions.}
\label{fig:jet-evolution}
\end{figure}

In Figure~\ref{fig:jet-evolution} we depict the two resummation schemes involved in the calculation of jet observables in heavy-ion collisions. Initially, the hard parton emerging from the hard QCD matrix element undergoes a DGLAP evolution from large angles $R_0 \sim 1$ down to the cone angle $R$. This is referred to as the $\log 1/R$ resummation \cite{Dasgupta:2014yra,Dasgupta:2016bnd,Kang:2016mcy,Dai:2016hzf}. The cross section for jet production in pp collisions therefore becomes
\beq
\label{eq:sigma-pp}
\sigma^{\tiny pp}(\pT,R) =  \sum_{k=q,g} f^{(n-1)}_{\jet/k} (R| \pT,R_0) \, \hat \sigma_k(\pT,R_0)\,,
\eeq
where $n\equiv n_k(\pT,R_0)$ is the power-index of the cross-section of the hard parton with flavor $k$. This is calculated at leading order (LO) at the factorization scale $Q^2_{\rm fac}$, such that $\hat \sigma_k = f_{i/p} \otimes f_{j/p} \otimes \hat \sigma_{ij \to k (l)} $, and involves a convolution of parton distribution functions (PDFs) $f_{i/p}(x,Q_\text{fac}^2)$ with the $2\to2$ QCD scattering cross section $\hat \sigma_{ij \to kl}$. The moment of the fragmentation function of an initial hard parton with flavor $k$ is $f^{(n)}_{\jet/k} (R| \pT,R_0) = \int_0^1 \dd x \, x^{n} f_{\text{jet}/k}(R|x,R_0)$, and receives both quark and gluon contributions, $f^{(n)}_{\jet/k} = \sum_{i=q,g}f^{(n)}_{i/k}$. In heavy-ion (AA) collisions, the baseline spectrum, referred to as $\tilde \sigma^{\tiny AA}(\pT,R)$, is modified by replacing the proton PDFs by nuclear PDFs $f_{i/p}(x,Q_\text{fac}^2) \to f_{i/A}(x,Q_\text{fac}^2)$.

The second stage of the evolution, see Fig.~\ref{fig:jet-evolution}, reflects the resummation of energy loss due to medium-induced processes. The full cross section in AA collisions therefore reads,
\begin{align}
\label{eq:sigma-AA}
\sigma^{\tiny AA}(\pT,R) \simeq \sum_{i=q,g} Q_i(\nu|\pT,R) \tilde \sigma^{\tiny AA}_i(\pT,R) \,, 
\end{align}
where $\tilde \sigma_{i}^{\tiny pp}$ is given in Eq.~\eqref{eq:sigma-pp}. Here, the \emph{quenching factor} of a jet $Q_i(\nu | \pT,R)$ accounts for the full energy loss of a jet including the partial recovery of ``lost'' energy within the jet cone. It depends on the Laplace variable $\nu= n_i(\pT)/\pT$, where $n_i(\pT)$ is the power of the steeply falling partonic spectrum, only through the initial conditions. The resummation of energy loss is calculated via a set of non-linear evolution equations \cite{Mehtar-Tani:2017web}, that read
\begin{align}
\label{eq:collimator-eq}
\frac{\del Q_i(p,\theta)}{\del \ln \theta} = \int_0^1 \dd z \,\frac{\alpha_s(k_\perp)}{2 \pi} p_{ji}^{(k)}(z) \, \Theta_\text{res}(z,\theta) \left[Q_j(zp,\theta) Q_k((1-z)p,\theta) - Q_i(p,\theta) \right] \,,
\end{align}
where $k_\perp = z(1-z) p \theta $, $p_{ji}^{(k)}(z)$ are the un-regularized Altarelli-Parisi splitting functions. The initial conditions are simply given by the partonic quenching factors for elastic and inelastic energy loss, i.e.  $Q_i(p,0) = Q^{(0)}_{\text{rad},i} (\nu) Q^{(0)}_{\text{el},i}(\nu) $, see \cite{Mehtar-Tani:2021fud} for further details.

The phase space constraint, encoded in $\Theta_{\rm res}$, ensures that only sufficiently hard jet splitting, i.e. those that form with short formation times inside of the medium, contribute to the total energy loss of the jet and is given by  $\Theta_\text{res}(z,\theta) = \theta(L-\tdecoh)\theta(\tdecoh -\tform )$.  The condition $\tform < \tdecoh$, where $\tform =2/[z(1-z)\pT\theta^2]$ and $\tdecoh = [12/(\hat q \theta^2)]^{1/3}$, restricts the early vacuum-like emissions to be harder than the medium scale. Finally, $\tdecoh < L$, or $\theta > \theta_c \equiv [12/(\hat qL^3)]^{1/2}$, makes sure that the jet splittings have sufficiently time to be resolved by medium interactions. This implies a minimal angle of the full resummation of jet energy loss as depicted by the red region in Fig.~\ref{fig:jet-evolution}.  Finally, in our semi-analytical treatment the medium parameters controlling elastic and inelastic energy loss, i.e. $\hat e = \hat q/(4T)$ and $\hat q$, are both sampled from a dynamically evolving hydrodynamical medium.  For further details, see \cite{Mehtar-Tani:2021fud}.

\section{Sensitivity to $\theta_c$ through path-length dependence}

\begin{figure}
\centering
\includegraphics[width=0.48\textwidth]{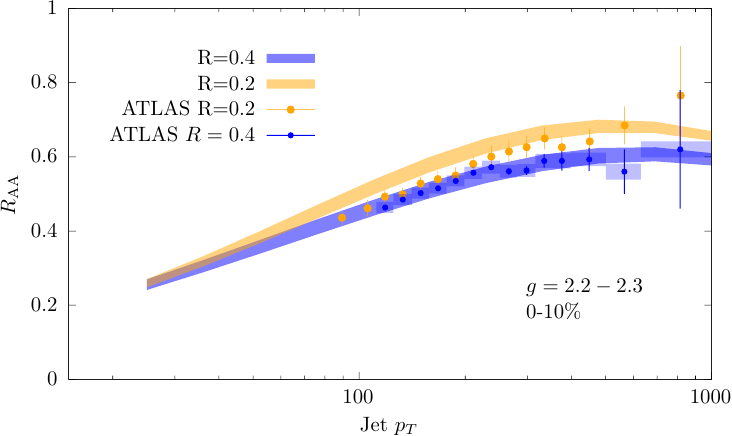}%
\hspace{0.5em}
\includegraphics[width=0.42\textwidth]{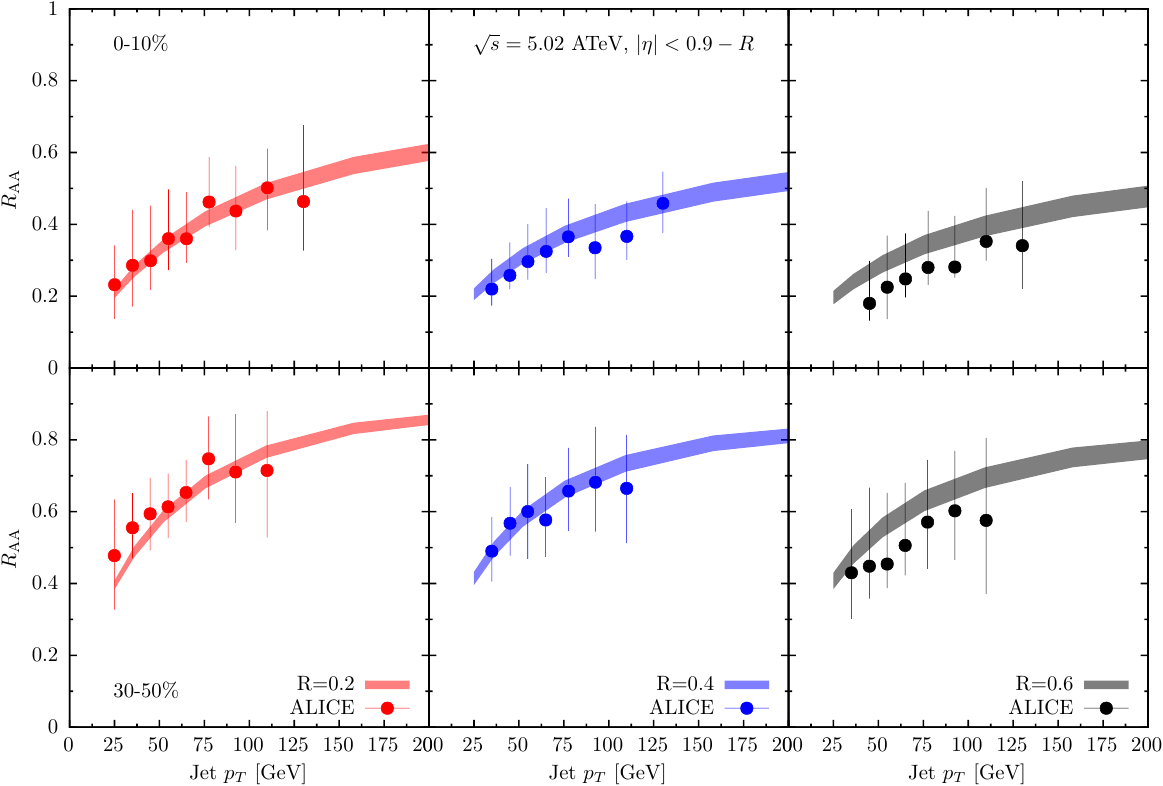}
\caption{Calculations of the nuclear modification factor for reconstructed jets for $R=0.2,\, 0.4$ and 0.6, see plots for details, compared to data from \cite{ATLAS:2023hso} (left) and \cite{ALICE:2023waz} (right).}
\label{fig:Raa-R}
\end{figure}

The description of experimental data over a wide $p_T$ and centrality range is remarkable \cite{Mehtar-Tani:2021fud}.\footnote{Note that after adjusting the two free parameters to mid-rapidity data \cite{Aaboud:2018twu}, no further tuning of the parameters was permitted.} The predicted $R$-dependence agrees well with the most recent experimental data \cite{CMS:2021vui,ATLAS:2023hso,ALICE:2023waz}.  As an example, we compare in Fig.~\ref{fig:Raa-R} our semi-analytical results with data from ATLAS and ALICE for central collisions heavy-ion collisions at $\sqrt{s_{\tiny NN}} = 5.02$ ATeV. Comparisons to data on reconstructed jets at RHIC energies are also in good agreement with recent measurements, and will be reported in a forthcoming publication.

From a thorough analysis of the associated theoretical uncertainties we found that the details of the phase space for in-medium vacuum-like emissions played a major role to describe the right jet suppression factor from small to moderate cone angles $R \leq 0.4$ \cite{Mehtar-Tani:2021fud}. Considering only gluon emission off a parton with color charge $C_R$, the phase space integral over the resolved phase space,
\beq
\Omega_{\rm res}(\pT,R) = \int_0^R \frac{\rmd \theta}{\theta} \int_0^1 \rmd z \, \frac{\alpha_s(k_\perp)}{2\pi} p_{gR}(z) \overset{{\rm DLA}}{\approx} \bar \alpha \left[\ln \frac{R}{\theta_c} \ln \frac{\pT}{\omega_c} + \frac23 \ln^2 \frac{R}{\omega_c}  \right]\,,
\eeq
where $\omega_c = \frac12 \hat q L^2$ is the critical energy and $\bar \alpha = \alpha_s C_R/\pi$. This result presumes that $R>\theta_c$ and $\pT > \omega_c$. Clearly, the role of both the energy of the jet as well as its cone angle contribute at single-logarithmic level.

To shed further light on the role of the phase space, we have calculated the jet production cross section as a function of azimuthal angle, i.e. 
\begin{equation}
    \frac{\rmd N^{\rm jet}}{\rmd \phi \,\rmd p_T}=\frac{N^{\rm jet}}{2 \pi}\left[1+2\sum_n v_n(p_T)\cos(n(\phi-\Psi_R))\right] \, .
\end{equation}
In particular, we focus on the second harmonic coefficient $v_2(\pT,R)$ which captures the path-length differences in- and out-of-plane due to the initial almond-shaped density distribution. The sensitivity to path-length becomes apparent if we approximate the flow coefficient by its linearized version, and write $v_2 \simeq [R_{\tiny AA}^{\rm in} - R_{\tiny AA}^{\rm out}]/[R_{\tiny AA}^{\rm in} + R_{\tiny AA}^{\rm out}]$, where $R_{\tiny AA}^{\rm in} = R_{\tiny AA}(L)$ and  $R_{\tiny AA}^{\rm out} = R_{\tiny AA}(L+\Delta L)$. Due to the many ways the path length enters the jet spectrum in Eq.~\eqref{eq:sigma-AA}, the resulting behavior is a complicated interplay between a wide variety of factors such as modifications of the quark/gluon fractions, sensitivity to path-length in the resolved phase space, modulations of the recovered energy within the jet cone etc. Currently, it is hard to pinpoint one general effect that stands out across the widely varying experimental conditions for measuring the jets.

\begin{figure}
\centering
\includegraphics[width=0.85\textwidth]{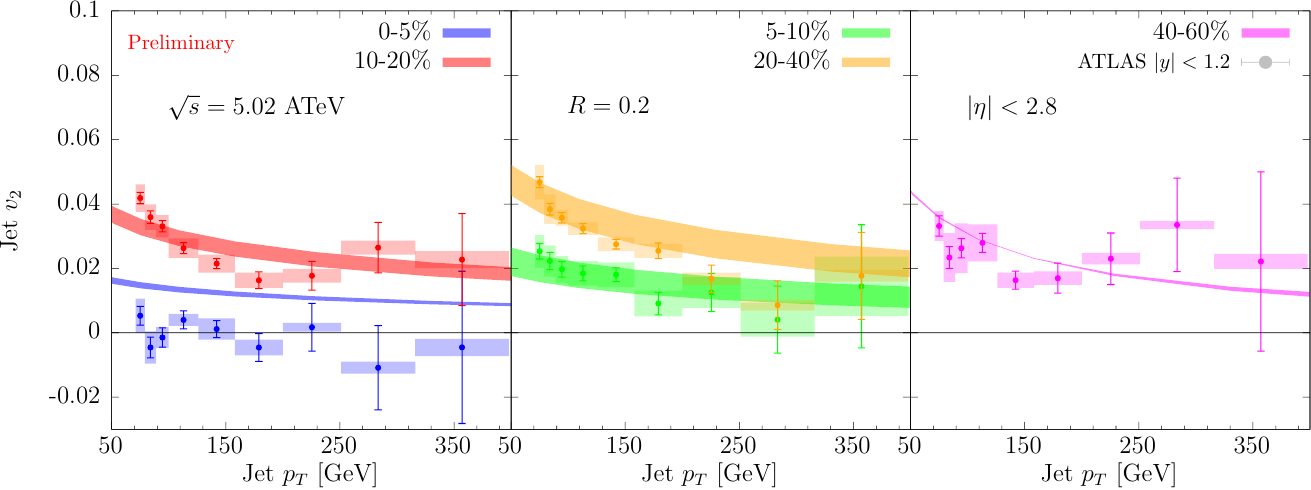}
\caption{Predictions for $v_2(p_T)$ for $R=0.2$ jets calculated for several centralities.}
\label{fig:v2-R}
\end{figure}

Our new calculation for these proceedings follows the same semi-analytic methodology as in the previous Section, except that we additionally bin the sampled jets in azimuthal angle to extract the $v_2$ coefficient. Our calculations are in excellent agreement with experimental data for reconstructed $R=0.2$ jets in a wide range of centralities \cite{ATLAS:2021ktw}, see Fig.~\ref{fig:v2-R}. We plan to present our predictions for the fully $R$- and $\pT$-dependent $v_2$ coefficient for jets for both RHIC and LHC kinematics in a forthcoming publication. We believe this novel observable to be promising to pin down the exact interplay between perturbative QCD evolution early in the medium, controlled by the evolution equations for the resummed quenching factors \eqref{eq:collimator-eq}, and the non-perturbative physics governing the distribution of soft fragments around the jet axis, resulting in the effective energy-loss that drives both $R_{\tiny AA}$ and $v_2$ at high-$\pT$.

\acknowledgments
Y. M.-T.'s work has been supported by the U.S. Department of Energy under Contract No. DE-SC0012704. D.P. has received funding from the European Union's Horizon 2020 research and innovation program under the Marie Sklodowska-Curie grant agreement No. 754496.

\bibliographystyle{JHEP}
\bibliography{hp2023}

\end{document}